# Title: Atomically thin mirrors made of monolayer semiconductors

**Authors:** Giovanni Scuri[1†], You Zhou[1,2†], Alexander A. High[1,2†], Dominik S. Wild[1†], Chi Shu[1], Kristiaan De Greve[1,2], Luis A. Jauregui[1], Takashi Taniguchi[3], Kenji Watanabe[3], Philip Kim[1*], Mikhail D. Lukin[1*] & Hongkun Park[1,2*]

**Affiliations:**

[1]Department of Physics and [2]Department of Chemistry and Chemical Biology, Harvard University, Cambridge, MA 02138, USA

[3]National Institute for Materials Science, 1-1 Namiki, Tsukuba 305-0044, Japan

†These authors contributed equally to this work.

*To whom correspondence should be addressed: hongkun_park@harvard.edu, lukin@physics.harvard.edu, pkim@physics.harvard.edu

**Abstract:** Transition metal dichalcogenide monolayers are promising candidates for exploring new electronic and optical phenomena and for realizing atomically thin optoelectronic devices. They host tightly bound electron-hole pairs (excitons) that can be efficiently excited by resonant light fields. Here, we demonstrate that a single monolayer of molybdenum diselenide ($MoSe_2$) can dramatically modify light transmission near the excitonic resonance, acting as an electrically switchable mirror that reflects up to 85% of incident light at cryogenic temperatures. This high reflectance is a direct consequence of the excellent coherence properties of excitons in this atomically thin semiconductor, encapsulated by hexagonal boron nitride. Furthermore, we show that the $MoSe_2$ monolayer exhibits power- and wavelength-dependent nonlinearities that stem from exciton-based

**lattice heating in the case of continuous-wave excitation and exciton-exciton interactions when fast, pulsed laser excitation is used. These observations open up new possibilities for studying quantum nonlinear optical phenomena and topological photonics, and for miniaturizing optical devices.**

**One Sentence Summary:** A molybdenum diselenide monolayer can act as an electrically switchable highly nonlinear resonant mirror with up to 85% reflectance at cryogenic temperatures.

**Main Text:** Mirrors are ubiquitous elements of optical and optoelectronic circuits. Their miniaturization is fundamentally limited by the optical wavelength in the case of dielectric mirrors and photonic crystals (*1*) or by the skin depth for metallic mirrors (*2*). Recently, resonant scattering has emerged as a method for overcoming these limitations and for controlling light at the atomic scale (*3-11*). For instance, highly reflective mirrors based on individual quantum emitters have been demonstrated by coupling them to optical cavities and nanophotonic waveguides (*3-8*). Such resonant mirrors feature very unusual properties due to their extraordinary nonlinearity down to the single-photon level (*3-8*). A two-dimensional (2D) layer of emitters, such as atomic lattices or excitons (*9-11*), has also been predicted to act as an efficient mirror when the incident light is resonant with the resonance frequency of the system. Such atomically thin mirrors represent the ultimate miniaturization limit of a reflective surface, and could enable unique applications ranging from quantum nonlinear optics (*9-11*) to topological photonics (*12, 13*).

This Report demonstrates that transition metal dichalcogenide (TMD) monolayers can act as atomically thin, electrically switchable, resonant mirrors. These materials are direct-bandgap semiconductors that support tightly bound excitons. Excitonic transitions in TMD monolayers exhibit large oscillator strengths (*14-16*), resulting in large radiative linewidths compared to excitons in other semiconductor systems. In addition, the excitonic response in monolayers can be controlled electrically via gate-induced doping and by shifting the chemical potential (*17-19*). Importantly, these monolayers can be easily integrated with other 2D materials via Van der Waals stacking to improve their quality or add new functionalities. One of the most studied amongst such heterostructures is a TMD monolayer encapsulated by two hexagonal boron nitride (hBN) flakes: this "passivated" monolayer exhibits enhanced carrier mobility (*19, 20*) and reduced photoluminescence linewidth (*21, 22*).

Our experiments make use of a device that consists of an hBN-passivated molybdenum diselenide ($MoSe_2$) monolayer placed on an oxide-covered silicon (Si) substrate, and we measure its reflectivity with a normally incident laser beam (Figs. 1A and 1B). The doped Si substrate is used as a gate electrode: by applying a gate voltage ($V_g$), $MoSe_2$ monolayers can be made intrinsic or *n*-doped. When a monochromatic laser beam is tuned to the exciton resonance, we observe substantial reflection from a monolayer device (**M1**) at $V_g < 10$ V at $T = 4$ K (Fig. 1C). The reflection contrast between the monolayer region and the substrate disappears at $V_g > 20$ V (Fig. 1D), indicating that the reflection can be turned off electrically. When we illuminate another monolayer device **(M2)** with a supercontinuum laser and spectrally resolve the reflection, we find that both the magnitudes and wavelength positions of the reflectance peaks change with $V_g$ (Fig. 1E). When the monolayer is intrinsic ($V_g < 10$ V), the reflection is dominated by a peak at the wavelength of the neutral exciton transition. When $MoSe_2$ is *n*-doped ($V_g > 20$ V), however, the reflection by the neutral exciton disappears, and a new, weaker, reflectance peak appears at the wavelength corresponding to the charged exciton resonance (*17-19*). These observations are consistent with previous studies (*17, 18*) that showed the disappearance (appearance) of the neutral (charged) exciton absorption in the *n*-doped region.

Figure 2A shows temperature-dependent reflectance spectra. At $T = 4$ K, the peak reflectance value reaches 0.8, demonstrating that the device can act as an efficient resonant mirror. Interestingly, the reflectance lineshape is not Lorentzian but rather displays a Fano-like asymmetry, exhibiting both a minimum and maximum near the neutral exciton resonance. As the temperature increases, the peak reflectance decreases while the linewidth increases. We find that the peak reflectance value varies in different samples or even within the same monolayer at distinct spatial locations, typically ranging from ~0.5 to 0.8. Some of these variations may stem from the intrinsic heterogeneity of the material (e.g. defects in the crystal) or charge/strain inhomogeneity introduced during the fabrication process.

In order to gain insight into our observations in Fig. 2A, we first consider the response of a free-standing MoSe$_2$ monolayer to normally incident coherent light, tuned close to the exciton resonance. On resonance, the optical field generated by the excitons interferes constructively (destructively) with the reflected (transmitted) light, resulting in enhanced reflection (suppressed transmission) (*11*). Because only a single exciton resonance is relevant over the frequency range of interest, the reflectance of a MoSe$_2$ monolayer can be described by the Lorentzian function (*10, 23*)

$$R(\omega) = \frac{\gamma_r^2}{2\tilde{\gamma}\gamma_T} \frac{1}{(\omega-\omega_0)^2/\tilde{\gamma}^2 + 1},\qquad(1)$$

where $\omega_0$ is the exciton resonance (angular) frequency, $\gamma_r$, $\gamma_T$ and $\tilde{\gamma}$ are the radiative decay rate, the total decay rate including non-radiative channels ($\gamma_T = \gamma_r + \gamma_{nr}$), and the dephasing rate, respectively. The dephasing rate $\tilde{\gamma}$ includes contributions from pure dephasing ($\gamma_d$) as well as from exciton decay: $\tilde{\gamma} = \gamma_d + \gamma_T/2$. The reflectance reaches a maximum at the resonance frequency $\omega_0$ with a peak value of $R_{\max} = (\gamma_r/\gamma_T)^2/(1+2\gamma_d/\gamma_T)$. In the absence of pure dephasing, non-radiative losses and disorder-induced broadening (*24*), this ratio is unity, and the MoSe$_2$ monolayer acts as a perfect mirror.

The presence of proximal dielectrics and other reflectors considerably modifies the optical response. Specifically, the lineshape shown in Fig. 2A originates from the interference between the MoSe$_2$ and substrate reflections. In our device, the MoSe$_2$ monolayer is embedded in a dielectric stack of hBN, SiO$_2$, and Si that acts as a multilayer broadband reflector (inset of Fig. 2A). Because the phase of the light reflected by the MoSe$_2$ monolayer varies by $\pi$ across the resonance, the interference between MoSe$_2$ reflection and substrate reflection produces an asymmetric Fano-like lineshape (Figure 2A). As such, the entire hBN/MoSe$_2$/hBN/SiO$_2$/Si heterostructure forms an effective cavity, which confines the optical field between the resonant MoSe$_2$ mirror and a broadband reflector, which can strongly modify the radiative decay of the excitonic system compared to its value in free space (*25*). As discussed in (*24*), for the present devices the decay rate is reduced by a factor of almost 3. This effect can viewed as resulting from the destructive interference of the emitted photon with itself upon reflecting off the substrate (*26-28*).

These effects are analyzed quantitatively using a master equation approach, that takes into account multiple reflections between the monolayer and the substrate using a transfer matrix method (*24, 29*). The results of this analysis are in excellent agreement with experimental observations (Fig. 2B and Fig. S1). By fitting the experimental spectra in Fig. 2A, we can determine the reflectance of the MoSe$_2$ monolayer itself and extract the critical parameters that characterize its optical properties, including radiative and non-radiative decay, and dephasing rates. Figure 2B shows the deconvolved reflectance spectra of the MoSe$_2$ monolayer in **M2** as a function of temperature. The peak reflectance at the exciton resonance reaches 0.85 at 4 K and remains larger than 0.8 below ~40 K (Fig. 2C). The total linewidth increases with increasing temperature (Fig. 2C). The extracted vacuum radiative linewidth ($\hbar\gamma_\mathrm{r}$) of the MoSe$_2$ monolayer is ~4.0 meV and is almost independent of temperature (Fig. 2D), in good agreement with recent experimental studies (*16, 30*). In contrast, both non-radiative decay and pure dephasing have a strong temperature dependence: at 4 K, pure dephasing is less prominent than non-radiative decay, and both are an order of magnitude smaller than the radiative linewidth. Their contributions to the linewidth become larger than the radiative linewidth as the temperature rises above 100 K.

We now turn to the behavior of our devices as a function of the incident laser power. When a continuous-wave (CW) laser is used as the excitation source (Fig. 3A), the reflectance of the device **M2** exhibits sudden jumps and prominent hysteresis as the laser is red-detuned from the exciton resonance. The threshold power at which the reflectance jump occurs becomes larger as the detuning is increased further to the red of the exciton resonance (Fig. 3A). When the wavelength of the CW laser beam is blue-detuned above the exciton resonance, the reflectance does not show any hysteresis, and, for high enough power, approaches the bare hBN-SiO$_2$-Si reflectance. Time-resolved measurements with an electro-optic modulator show that the reflectivity jumps occur on a nanosecond timescale (Fig. S2).

We also explore power-dependent nonlinearities using pulsed excitation (6-ps full width at half maximum duration). As shown in Fig. 4, the observed nonlinear response is dramatically different from that obtained using CW excitation. First, the reflectance jumps and hysteresis completely disappear with the picosecond laser excitation. Second, the power dependence of the Fano-like lineshape also changes. Specifically, when the wavelength of the picosecond laser is tuned to the blue side of the Fano-like dip (Fig. 4A), the **M2** reflectance exhibits an initial decrease followed by a 10-fold increase with increasing peak laser power ($I_p$). Figure 4B shows a two-dimensional plot of the device reflection as a function of incident wavelength and power, demonstrating a blueshift of the dip of Fano resonance. At very large $I_p$ (~1 W), the reflectance eventually converges to that of the hBN-SiO$_2$-Si reflectance (Fig. 4A).

The power-dependent changes in reflectance presented in Figs. 3 and 4 can be explained by considering laser-induced heating and exciton-exciton interactions, respectively. In particular, the reflectance jump and hysteresis observed in Fig. 3A can be understood in terms of the redshift of the exciton resonance resulting from laser-induced heating, Fig. 2B (*24*). As the CW laser power increases, more and more excitons are generated per unit time. Some of these decay through non-radiative channels, increasing the temperature of the $MoSe_2$ monolayer. When the wavelength of the laser is red detuned from the exciton resonance, this temperature increase brings the exciton resonance closer to the laser wavelength (Fig. 2B), giving rise to increased absorption. Such positive feedback leads to bistability in the reflectivity (Fig. 3B) and to the reflectance jump observed in Fig. 3A. Once the resonance is moved to match the excitation wavelength, it is locked in its energy even at weaker excitation power, leading to hysteresis. As shown in Fig. 3C, this simple model, in combination with the temperature-induced resonance shift in Fig. 2B and non-radiative rates determined in Fig. 2D, qualitatively explains the experimental data. We note that this model does not take into account the temperature dependence of the heat capacity or thermal conductivity of $MoSe_2$, which may contribute to some discrepancies with the data. The ~100 K increase in temperature required to explain the experimentally observed redshift in Fig. 3A is consistent with a finite element simulation of our device temperature with the relevant parameters (Fig. S3).

The power-dependence of the reflection in the picosecond regime (Fig. 4) cannot be explained by laser-induced heating. First, as shown in Fig. 4B, the dip in the Fano-like lineshape blueshifts as the laser power increases, in clear contrast to the temperature-induced redshift observed in Fig. 2A. Second, the time scale of the laser pulse (6 ps) during which this blueshift happens is much faster than the time scale associated with the thermally-induced process in Fig. 3 (~1 ns), while the time interval between the pulses (12.5 ns) is much longer.

Instead, this ultra-fast nonlinearity is likely associated with exciton-exciton interactions. These can give rise to a density-dependent blueshift as well as collisional broadening of the exciton resonance (*31*). We model these exciton interaction effects by introducing the density-dependent blueshift of the exciton resonance $\Delta(n) = an$, the change in the dephasing rate $\delta\gamma_d(n) = bn$, and the change in the non-radiative decay rate $\delta\gamma_{nr}(n) = cn$, where $n$ denotes the exciton density (*24*). By fitting this model to experimental data, we extract the three parameters related to exciton interactions: $\hbar a = 9.6\times10^{-13}$ meV·cm$^2$, $\hbar b = 9.4\times10^{-13}$ meV·cm$^2$, and $\hbar c = 4.0\times10^{-13}$ meV·cm$^2$. These values indicate that an energy shift equal to the radiative linewidth occurs when excitons are separated by ~8 nm. The results of this analysis, displayed in Fig. 4C, are in good agreement with experimental observations. The observed shifts are consistent with the theoretical estimates for exchange interactions, $a_{\text{thoery}} \sim E_B R^2 \sim 5\times10^{-12}$ meV·cm$^2$ (*32*), where $E_B \approx 500$ meV is the exciton binding energy and $R$ (~1 nm) denotes the exciton Bohr radius. Detailed understanding of the observed nonlinearities, including the interplay between shifts and dephasing requires additional theoretical and experimental studies.

Our observations open up intriguing prospects for exploring novel phenomena and device applications based on TMD excitons. In particular, our observations demonstrate the remarkable optical quality of hBN-passivated MoSe$_2$ monolayers, as indicated by the radiative decay rate being more than 20 times larger than the non-radiative and pure dephasing decay rates (Fig. 2D). These properties enable detailed investigation of fundamental exciton physics, as well as unique applications in quantum optics. Such applications may include engineering of robust, long-lived optical edge states using periodically modulated Moiré heterostructures in external magnetic field (*12, 13*), controlling the emission patterns of localized sources using atomically thin metasurfaces (*10*) and realization of quantum nonlinear optical systems featuring broadband optical squeezing (*11*) or strong interaction between photons (*10*). Finally, engineering of the exciton optical response using broad-band reflectors (Figure 2A) or other properly designed photonic systems can be explored for further enhancement of the nonlinearities (*24*). Technologically, atomically thin mirrors also have intriguing potential. The electrical tunability can enable its use as a reconfigurable component in a wide variety of optical and optoelectronic systems including modulators, active cavities, and active metasurfaces. Strong, fast, intrinsic and engineered nonlinearities make these systems excellent candidates for realizing optoelectronic devices with possible applications in classical and quantum information processing.

**References and Notes:**

1. J. D. Joannopoulos, S. G. Johnson, J. N. Winn, R. D. Meade, *Photonic Crystals: Molding the Flow of Light*. (Princeton University Press, 2011).

2. B. E. Saleh, M. C. Teich, B. E. Saleh, *Fundamentals of Photonics*. (Wiley New York, 1991), vol. 22.

3. D. Englund *et al.*, Controlling cavity reflectivity with a single quantum dot. *Nature* **450**, 857-861 (2007).

4. D. E. Chang, A. S. Sørensen, E. A. Demler, M. D. Lukin, A single-photon transistor using nanoscale surface plasmons. *Nature Phys.* **3**, 807-812 (2007).

5. K. Hennessy *et al.*, Quantum nature of a strongly coupled single quantum dot-cavity system. *Nature* **445**, 896-899 (2007).

6. D. O'Shea, C. Junge, J. Volz, A. Rauschenbeutel, Fiber-optical switch controlled by a single atom. *Phys. Rev. Lett.* **111**, 193601 (2013).

7. J. Thompson *et al.*, Coupling a single trapped atom to a nanoscale optical cavity. *Science* **340**, 1202-1205 (2013).

8. A. Sipahigil *et al.*, An integrated diamond nanophotonics platform for quantum-optical networks. *Science* **354**, 847-850 (2016).

9. R. J. Bettles, S. A. Gardiner, C. S. Adams, Enhanced optical cross section via collective coupling of atomic dipoles in a 2D array. *Phys. Rev. Lett.* **116**, 103602 (2016).

10. E. Shahmoon, D. S. Wild, M. D. Lukin, S. F. Yelin, Cooperative resonances in light scattering from two-dimensional atomic arrays. *Phys. Rev. Lett.* **118**, 113601 (2017).

11. S. Zeytinoglu, C. Roth, S. Huber, A. Imamoglu, Atomically thin semiconductors as nonlinear mirrors. *arXiv:1701.08228*, (2017).

12. F. Wu, T. Lovorn, A. H. MacDonald, Topological cxciton bands in Moiré heterojunctions. *Phys. Rev. Lett.* **118**, 147401 (2017).

13. J. Perczel *et al.*, Topological Quantum Optics in Two-Dimensional Atomic Arrays. *arXiv:1703.04849*, (2017).

14. K. F. Mak, C. Lee, J. Hone, J. Shan, T. F. Heinz, Atomically thin $MoS_2$: A new direct-gap semiconductor. *Phys. Rev. Lett.* **105**, 136805 (2010).

15. A. Splendiani *et al.*, Emerging photoluminescence in monolayer $MoS_2$. *Nano Lett.* **10**, 1271-1275 (2010).

16. G. Moody *et al.*, Intrinsic homogeneous linewidth and broadening mechanisms of excitons in monolayer transition metal dichalcogenides. *Nat. Commun.* **6**, 8315 (2015).

17. K. F. Mak *et al.*, Tightly bound trions in monolayer $MoS_2$. *Nature Mater.* **12**, 207-211 (2013).

18. B. Ganchev, N. Drummond, I. Aleiner, V. Fal'ko, Three-particle complexes in two-dimensional semiconductors. *Phys. Rev. Lett.* **114**, 107401 (2015).

19. M. Sidler *et al.*, Fermi polaron-polaritons in charge-tunable atomically thin semiconductors. *Nature Phys.* **13**, 255-261 (2017).

20. C. R. Dean *et al.*, Boron nitride substrates for high-quality graphene electronics. *Nature Nanotech.* **5**, 722-726 (2010).

21. Y. Zhou *et al.*, Probing dark excitons in atomically thin semiconductors via near-field coupling to surface plasmon polaritons. *arXiv:1701.05938*, (2017).

22. O. A. Ajayi *et al.*, Approaching the Intrinsic Photoluminescence Linewidth in Transition Metal Dichalcogenide Monolayers. *arXiv:1702.05857*, (2017).

23. M. M. Glazov *et al.*, Exciton fine structure and spin decoherence in monolayers of transition metal dichalcogenides. *Phys. Rev. B* **89**, 201302 (2014).

24. Materials and methods as well as supplementary online information are available as supplementary materials.

25. E. M. Purcell, H. C. Torrey, R. V. Pound, Resonance absorption by nuclear magnetic moments in a solid. *Phys. Rev.* **69**, 37-38 (1946).

26. C. Weisbuch, M. Nishioka, A. Ishikawa, Y. Arakawa, Observation of the coupled exciton-photon mode splitting in a semiconductor quantum microcavity. *Phys. Rev. Lett.* **69**, 3314-3317 (1992).

27. M. Hübner *et al.*, Collective effects of excitons in multiple-quantum-well Bragg and anti-Bragg structures. *Phys. Rev. Lett.* **76**, 4199-4202 (1996).

28. S. Haas *et al.*, Intensity dependence of superradiant emission from radiatively coupled excitons in multiple-quantum-well Bragg structures. *Phys. Rev. B* **57**, 14860-14868 (1998).

29. H. A. Macleod, *Thin-Film Optical Filters*. (CRC press, 2001).

30. M. Selig *et al.*, Excitonic linewidth and coherence lifetime in monolayer transition metal dichalcogenides. *Nat. Commun.* **7**, (2016).

31. C. Ciuti, V. Savona, C. Piermarocchi, A. Quattropani, P. Schwendimann, Role of the exchange of carriers in elastic exciton-exciton scattering in quantum wells. *Phys. Rev. B* **58**, 7926-7933 (1998).

32. G. Rochat *et al.*, Excitonic Bloch equations for a two-dimensional system of interacting excitons. *Phys. Rev. B* **61**, 13856-13862 (2000).

33. P. J. Zomer, M. H. D. Guimarães, J. C. Brant, N. Tombros, B. J. van Wees, Fast pick up technique for high quality heterostructures of bilayer graphene and hexagonal boron nitride. *Appl. Phys. Lett.* **105**, 013101 (2014).

34. K. S. Gavrichev *et al.*, Low-temperature heat capacity and thermodynamic properties of four boron nitride modifications. *Thermochim. Acta* **217**, 77-89 (1993).

35. E. K. Sichel, R. E. Miller, M. S. Abrahams, C. J. Buiocchi, Heat capacity and thermal conductivity of hexagonal pyrolytic boron nitride. *Phys. Rev. B* **13**, 4607-4611 (1976).

**Acknowledgments**

We acknowledge support from the DoD Vannevar Bush Faculty Fellowship (N00014-16-1-2825), AFOSR MURI (FA9550-12-1-0024 and FA9550-17-1-0002), NSF (PHY-1506284), NSF CUA (PHY-1125846), the Gordon and Betty Moore Foundation, and Samsung Electronics. We carried out all film deposition and device fabrication at the Harvard Center for Nanoscale Systems.

**Supplementary Materials:**

Materials and Methods

Figures S1-S3

References (33-35)

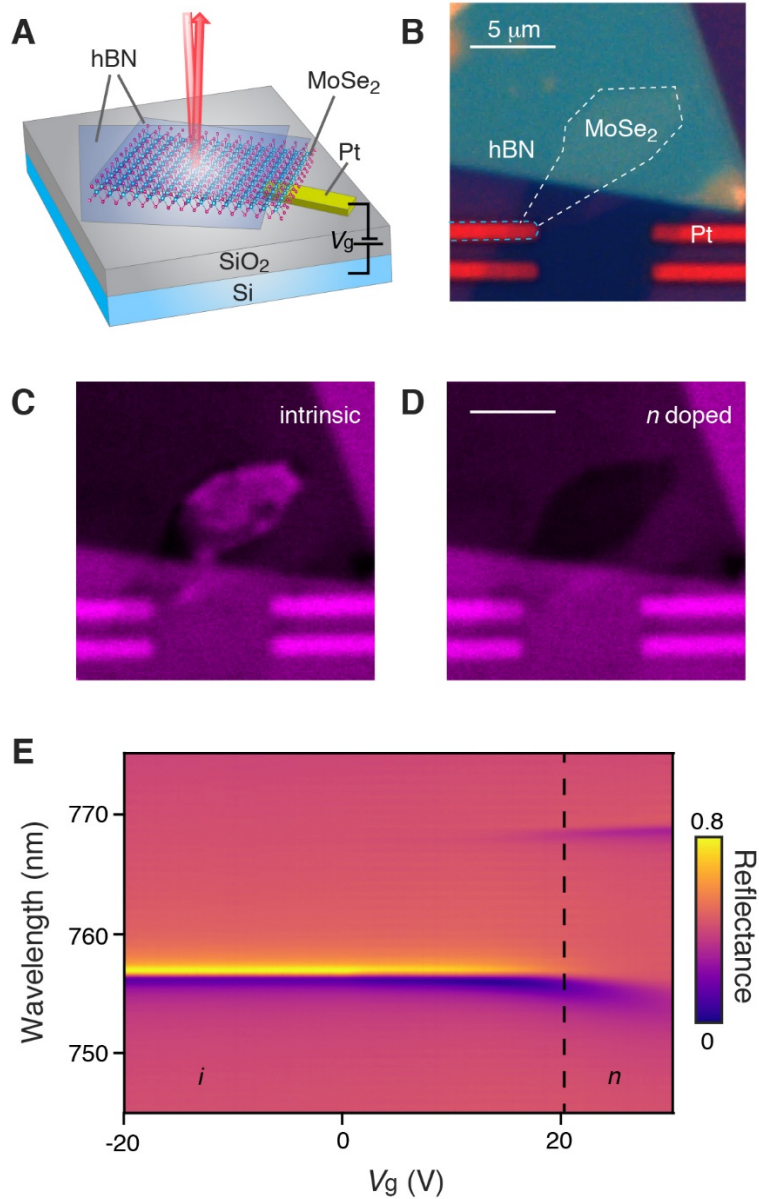

**Fig. 1**. **Electrically switchable, atomically thin mirror.** (**A**) Schematic of the experimental setup. A MoSe$_2$ monolayer encapsulated between two hBN layers is placed on SiO$_2$ (285 nm)/Si substrates. Platinum (Pt) electrodes are used to contact the MoSe$_2$ monolayer, while Si is used as back gate. The incident light resonant with the exciton transitions in MoSe$_2$ can be strongly reflected. The free carrier density in MoSe$_2$ can be tuned with a gate voltage applied to the Si back gate. (**B**) Optical image of a monolayer MoSe$_2$ device **M1**. The dashed white and blue lines

show the outline of MoSe$_2$ and a Pt contact, respectively. **(C)** and **(D)** Reflection images of **M1** under 750-nm resonant laser illumination with a gate voltage of **(C)** = -20 V and **(D)** 30 V at $T =$ 4 K. The fully encapsulated MoSe$_2$ region shows substantial reflection, while the part that is not fully encapsulated is not as reflective. **(E)** Reflection spectra of another monolayer MoSe$_2$ device **M2** as a function of gate bias at 4 K. Reflection due to charged excitons can be observed in the electron-doped regime. To obtain absolute reflectance, we normalize the reflected intensity using the measured reflectance of the bare substrate and metal electrodes.

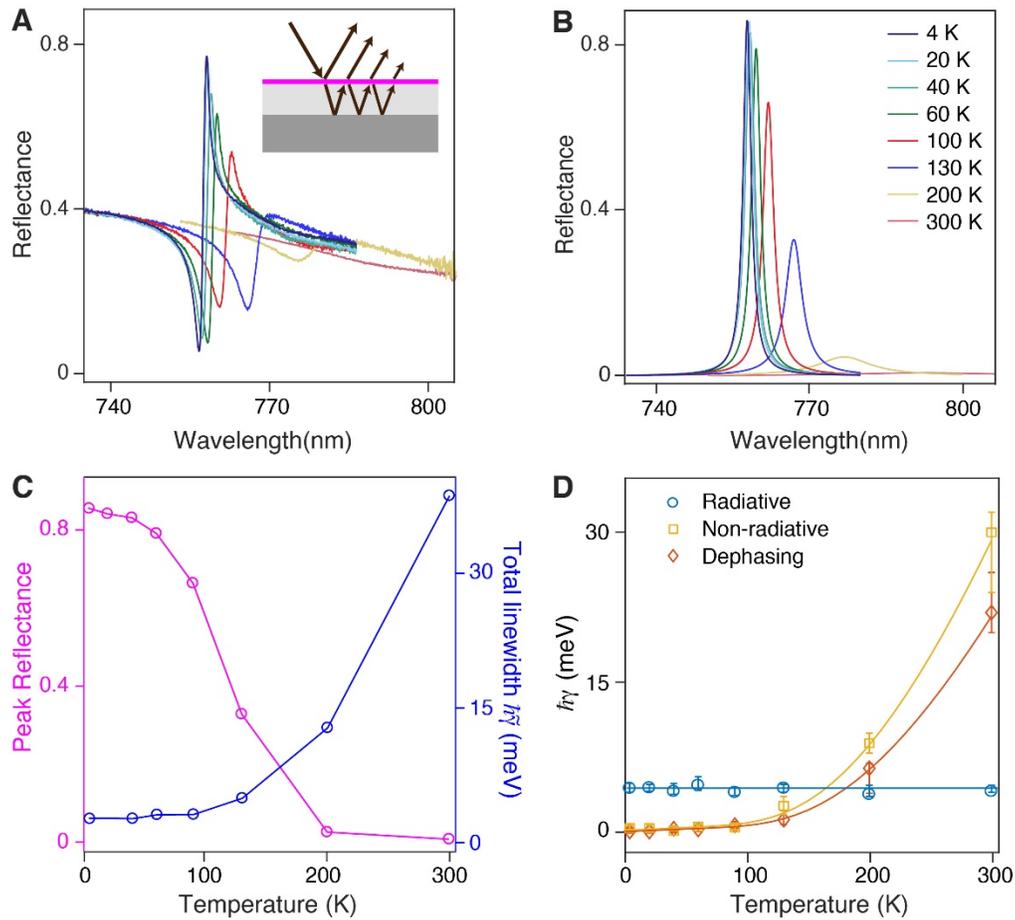

**Fig. 2. Temperature-dependent reflectance spectra of a MoSe$_2$ monolayer.** **(A)** Temperature dependent reflection spectra of the monolayer MoSe$_2$ device **M2**. The line colors correspond to different temperatures listed in (**B**). **(B)** Reflectance spectra of the MoSe$_2$ monolayer as a function of temperature obtained from the fit. Inset: schematic of interference between the reflection from the MoSe$_2$ monolayer and the reflection of the BN/SiO2/Si substrate **(C)** Temperature dependence of peak reflectance and total linewidth of the MoSe$_2$ monolayer in **M2**. **(D)** Radiative, non-radiative and pure dephasing linewidths of the same monolayer as a function of temperature.

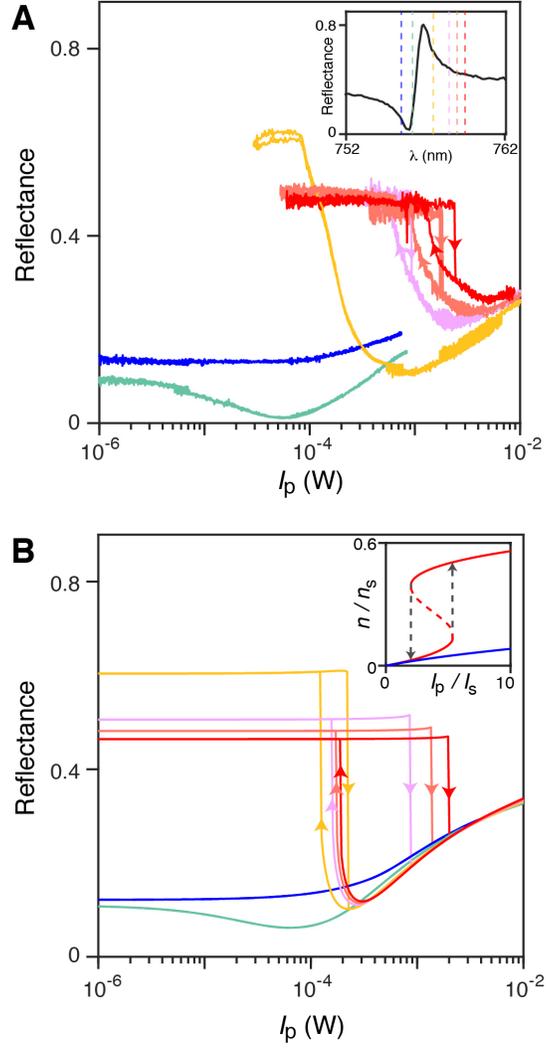

**Fig. 3. Nonlinearity of the reflection from a MoSe$_2$ monolayer under continuous-wave (CW) resonant excitation.** **(A)** Reflectance of the monolayer MoSe$_2$ device **M2** as a function of the CW laser power ($I_p$) at different detuning wavelengths, as indicated in the inset. **(B)** Nonlinear reflectance predicted by the thermal model (*24*) at different detuning wavelengths. Solid and dashed line denotes increasing and decreasing power, respectively. Inset: schematic of the optical bistability with red detuning. Here, $n_s$ and $I_s$ denote the photon density and the incident laser peak power when the energy shift of the exciton transition is equal to the radiative linewidth.

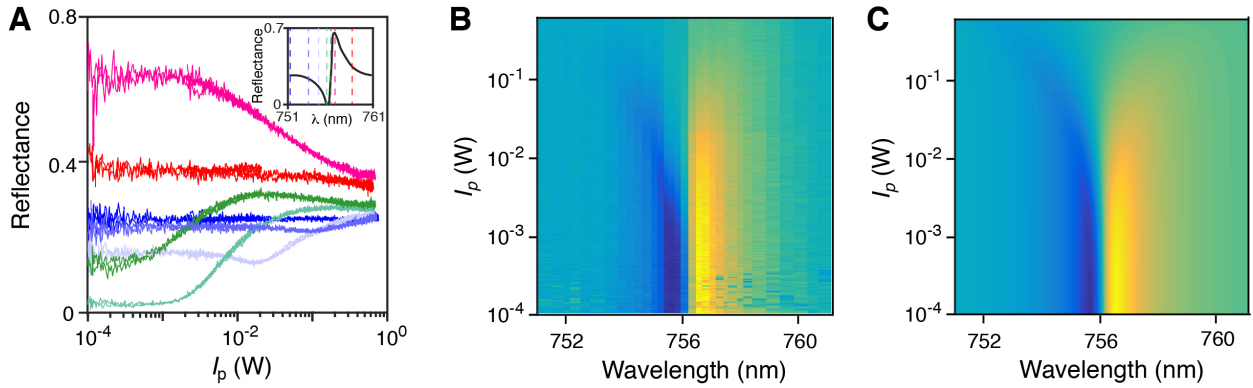

**Fig. 4. Nonlinearity of the optical response under pulsed-laser excitation. (A)** Reflection as a function of peak power ($I_p$) of the picosecond laser excitation (6ps full width at half maximum), at the wavelengths indicated in the inset. **(B) and (C)** 2D plots of **(B)** experimental and **(C)** theoretical reflectance at various excitation wavelengths as a function of laser peak power.

**Materials and Methods**

Thin hBN layers (10 – 180-nm thick) and monolayer $MoSe_2$ were mechanically exfoliated from bulk crystals onto $SiO_2$(285 nm)/Si substrates, and were identified based on their optical contrasts. The thickness of the selected hBN flakes was determined by atomic force microscopy measurements. Prior to exfoliation, the $SiO_2$/Si substrates were first ultrasonically cleaned in acetone and 2-propanol for 5 minutes, and then with a piranha solution (2:1 96% concentrated sulfuric acid/30% hydrogen peroxide) for 5 minutes. Following the solution-cleaning step, the substrates were subjected to further cleaning in oxygen plasma (100 W, 300 mTorr) for 10 minutes. Afterwards, the flakes were exfoliated by the scotch-tape method. The hBN/$MoSe_2$/hBN heterostructures were then stacked and transferred using a dry transfer method (33) onto cleaned $SiO_2$/Si substrates. Electrical contacts patterned by electron-beam lithography were deposited by electron-beam evaporation of Cr (10 nm)/Au (90 nm). Optical measurements were carried out in a home-built confocal microscope using an objective with a numerical aperture of 0.75 in a cryostat from Montana Instruments. For resonant reflection measurements, Ti:sapphire lasers in the continuous-wave (CW) mode (M Squared) and pulsed mode (Coherent Mira) were used for excitation. To obtain the reflectance spectra, we either scanned the CW Ti:sapphire laser across the wavelength range or used a SuperK supercontinuum laser with a broad spectral range.

**Supplementary Text**

Reflection of TMD monolayers

We consider a normally incident plane wave reflecting off a MoSe$_2$ monolayer, which is embedded between an arbitrary number of dielectric layers. We assume that the system is translationally invariant in the *xy* plane, with the incident plane wave propagating along the *z* axis. In order to determine the reflection coefficient including both dephasing and non-radiative decay of the excitons, we start from a fully quantized treatment of both the excitons and the electric field. Within the Born–Markov approximation, the positive frequency part of the electric field operator evolves according to *(10)*

$$E^+(z,t) = E_0^+(z,t) + \frac{k_0^2}{\varepsilon_0} G(z) P_{2D}^+(t), \tag{S1}$$

where $E_0^+(z,t)$ is the free evolution of the electric field (in the absence of the MoSe$_2$ monolayer), $k_0$ is the free space wavenumber of the incident field, $G(z)$ is the electromagnetic Green's function for the given medium evaluated at the exciton resonant frequency, and $P_{2D}^+(t)$ is the two-dimensional polarization of the TMD monolayer. The Green's function can be computed using the transfer matrix techniques discussed below. The polarization operator can be related to the exciton annihilation operator $c$ (at zero momentum for normal incidence) by $P_{2D}^+ = \wp\, c$, where $\wp$ is the dipole moment of the exciton transition.

At the position $z_0$ above the top dielectric layer, the reflected electric field is given by

$$E_r^+(t) = r_0 e^{ik_0 z_0} E_{in}^+(t) + \frac{k_0^2}{\varepsilon_0} G(z_0) \wp\, c(t), \tag{S2}$$

where $r_0$ denotes the complex reflection coefficient of the entire stack of dielectric layers. For a monochromatic input field in a coherent state, we obtain the reflectance

$$R = |r_0|^2 + \left(\frac{k_0 \wp}{2\varepsilon_0}\right)^2 |G(z_0)|^2 \frac{\langle c^\dagger c \rangle}{|E_{in}^+|^2} + 2\frac{k_0^2 \wp}{\varepsilon_0} \text{Re}\left[G(z_0)^* r_0 e^{ik_0 z_0} \frac{\langle c \rangle}{E_{in}^+}\right]. \tag{S3}$$

The first and second term describe the reflection by the substrate and the MoSe$_2$ monolayer, respectively, while the last term captures their interference. The expectation values $\langle c \rangle$ and $\langle c^\dagger c \rangle$ can be computed using a master equation approach. In particular, given a total dephasing rate $\tilde{\gamma}$ and a population decay rate $\gamma$, we have:

$$\frac{d}{dt}\langle c \rangle = i(\delta - \Delta + i\tilde{\gamma})\langle c \rangle + i\wp E_0^+$$
$$\frac{d}{dt}\langle c^\dagger c \rangle = -\gamma \langle c^\dagger c \rangle + i\wp E_0^+ \langle c^\dagger \rangle - i\wp E_0^- \langle c \rangle. \tag{S4}$$

Here, $\delta$ is the detuning from the exciton resonance in vacuum, and $\Delta = \gamma_r k_0 \text{Re}[G(0)]/2$ is a shift of the resonance due to the presence of the substrate. $\gamma_r$ denotes the radiative decay rate of the excitons in free space, which is given by

$$\gamma_r = \frac{k_0 \wp^2}{\varepsilon_0}. \tag{S5}$$

It is important to note that the radiative rate is modified by the substrate to become $2\gamma_r k_0 \text{Im}[G(0)]$. To simplify notation, we introduce the decay rates

$$\gamma = \gamma_{nr} + 2\gamma_r k_0 \text{Im}[G(0)], \quad \tilde{\gamma} = \gamma_d + \frac{1}{2}\gamma, \tag{S6}$$

where $\gamma_{nr}$ denotes the non-radiative decay rate, and $\gamma_d$ corresponds to pure dephasing. The reflectance may now be readily obtained by solving for the steady-state values of $\langle c \rangle$ and $\langle c^\dagger c \rangle$ in Eq. (S4) and substituting back into Eq. (S3). In the absence of a substrate, $r_0 = 0$ and $G(z) = \dfrac{i}{2k_0} e^{ik_0 z}$, we find

$$R = \frac{R_{max}}{\delta^2/\tilde{\gamma}^2 + 1}, \tag{S7}$$

where

$$R_{max} = \frac{(\gamma_r/\gamma)^2}{1 + 2\gamma_d/\gamma} = \frac{QE^2}{1 + 2\gamma_d/\gamma}. \tag{S8}$$

Therefore the peak reflection from a suspended TMD monolayer provides a lower bound on the quantum efficiency ($QE = \gamma_r/\gamma$) or, conversely, an upper bound on the dephasing rate.

Inhomogeneous broadening

In order to account for inhomogeneous broadening, we assume that the exciton resonant frequency follows a Gaussian distribution

$$p(\omega) = \frac{1}{\sqrt{2\pi\gamma_i^2}} e^{-(\omega-\omega_0)^2/2\gamma_i^2}, \tag{S9}$$

where $\gamma_i$ is the inhomogeneous linewidth. Provided that there is no interference between light emitted by excitons with different resonant frequencies, the observed reflectance will be given by the ensemble average, and the peak reflectance is given by

$$\bar{R}_{max} = \bar{R}(0) = \sqrt{\frac{\pi}{2}} \frac{\tilde{\gamma}}{\gamma_i} e^{\tilde{\gamma}^2/2\gamma_i^2} \operatorname{erfc}\left(\frac{\tilde{\gamma}}{\sqrt{2}\gamma_i}\right) R_{max} \leq R_{max}. \tag{S10}$$

Hence, this establishes a lower bound on the quantum efficiency in the presence of inhomogeneous broadening, or conversely and upper bound on the inhomogeneous linewidth.

Transfer matrix method

The reflectance of multi-layer thin films can be treated by the so-called transfer matrix method (*29*). In this method, forward- and backward-traveling waves form the basis into which the optical electric field in each layer is decomposed. The field at an arbitrary point in the layer can be computed by matrix multiplication, where each matrix describes the light propagation at an interface or inside a homogeneous material. In particular, under normal incidence condition, the transfer matrix for light propagating across the interface between two dielectrics (from $\varepsilon_1$ to $\varepsilon_2$, assuming magnetic susceptibility is one for both materials) is given by:

$$M(\varepsilon_1, \varepsilon_2) = \frac{1}{2\sqrt{\varepsilon_2}} \begin{pmatrix} \sqrt{\varepsilon_1} + \sqrt{\varepsilon_2} & -\sqrt{\varepsilon_1} + \sqrt{\varepsilon_2} \\ -\sqrt{\varepsilon_1} + \sqrt{\varepsilon_2} & \sqrt{\varepsilon_1} + \sqrt{\varepsilon_2} \end{pmatrix}. \quad (S11)$$

The transfer matrix for light propogation within a material of dielectric constant $\varepsilon$ is

$$M = \begin{pmatrix} e^{ikd} & 0 \\ 0 & e^{-ikd} \end{pmatrix}, \quad (S12)$$

where $k$ is wavevector inside the material determined by $\varepsilon$ and the vacuum wavevector, and $d$ is the thickness of the layer.

The reflection and transmission of the entire structure can be calculated by multiplication of all the transfer matrix for each layer and interface. Similarly, the Green's function that appears in Eq. (S1) is obtained by placing a (two-dimensional) dipole source at the location of the MoSe$_2$ monolayer and propagating the field outwards. By combining these results with the expression for the reflectance in Eq. (S3), we can thus compute the reflectance at a given frequency in terms of the four parameters $\omega_0$, $\gamma_r$, $\gamma_d$, and $\gamma_{nr}$ (Fig. S1A). For example, for an ideal monolayer of MoSe$_2$ ($\gamma_d = \gamma_{nr} = 0$) is placed above a simple substrate with reflection coefficient $r_0$, we obtain the reflectance

$$R = \left| r_0 - \frac{(1+r_0)^2}{1+\mathrm{Re}(r_0)} \frac{i[1+\mathrm{Re}(r_0)]\gamma_r/2}{\delta - \mathrm{Im}(r_0)\gamma_r/2 + i[1+\mathrm{Re}(r_0)]\gamma_r/2} \right|. \tag{S13}$$

The reflectance takes the shape of a broadband reflector ($r_0$) interfering with a Lorentzian resonance with linewidth $[1+\mathrm{Re}(\gamma_0)]\gamma_r$, which explains the Fano-like lineshape observed in experiment. We note that the width of the resonance can be considerably modified by the presence of the reflector compared to the width in free space, $\gamma_r$. In particular, the radiative linewidth can be strongly reduced if the monolayer is placed near the node of a highly reflecting substrate ($r_0 \sim 1$). We point out that in practice, the narrowest achievable resonance is bounded by non-radiative decay and dephasing, limiting the extent of modification to about $\gamma_r/20$ for the present samples.

We compare these theoretical predictions with experiment by fitting the unknown parameters. This analysis yields excellent agreement with the data (Fig. S1A). It further enables us to deconvolve the temperature-dependent reflection of the MoSe$_2$ monolayer from the temperature-independent reflection of the BN/SiO$_2$/Si layers (Fig. S1B). For the structures explored in the main text, which include a top layer of hBN in addition to the reflecting substrate, we find that the modified radiative width is given by approximately 0.37 $\gamma_r$. The free-space radiate linewidth $\gamma_r$ extracted using this analysis is consistent with values reported previously (*16, 30*).

We also note that the Fano-like line shape is affected differently by non-radiative decay and dephasing. Specifically, both non-radiative decay and dephasing rate can modify the asymmetry between the peak and dip in the Fano-like line in a different fashion, as shown in Fig. S1C. The fit is significantly improved when considering both contributions. We estimate the error bars of $\gamma_{nr}$ and $\gamma_d$ by their values at which the fitting error is twice as large as the minimum error at the best fit.

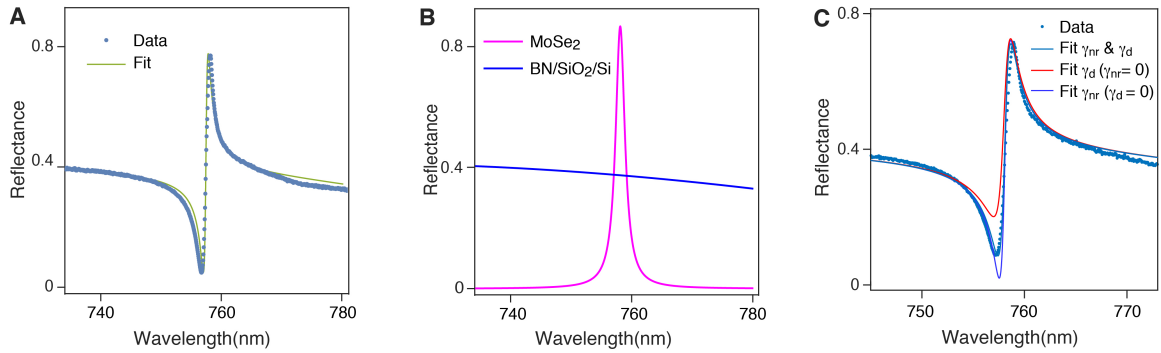

**Figure S1.** (**A**) Measured reflection spectra of device M2, $MoSe_2$ monolayer encapsulated in BN on $SiO_2$/Si substrate at 4 K. The line shows the fit by the model. (**B**) The reflection from $MoSe_2$ monolayer and BN/$SiO_2$/Si background after deconvolution. (**C**) Measured reflection spectra of device M2 at 40 K. Non-radiative and dephasing rate modifies the assymetry of the Fano-like line shape differently. The fit considering both nonradiative and dephasing matches well with the data.

Temperature dependence of nonradiative decay and dephasing rate

The temperature-dependence of non-radiative decay $\gamma_{nr}$ and pure dephasing $\gamma_d$ rate can be phenomenologically fitted to

$$\gamma_i = \gamma_{1,i} + c_{ac,i} T + \frac{c_{LO,i}}{e^{\frac{\Omega}{kT}} - 1}, \qquad \text{(S14)}$$

where $i$ can be non-radiative rate or pure dephasing, and the first term $\gamma_1$ corresponds to the residual broadening at zero temperature, while the second and third terms describe the broadening due to coupling between excitons and acoustic and longitudinal optical (LO) phonons, respectively. In Equation (S14), $c_{ac}$ and $c_{LO}$ represent the exciton-phonon coupling strengths for acoustic and LO phonons, respectively, and $\Omega$ is the average energy of phonons in the LO branch. The fitting is presented as solid line in Fig. 2D.

Analysis of non-linear response

The model presented above may be extended to include nonlinear effects. In order to obtain the power-dependent reflectance in Fig. 3C, the non-linearity instead enters via the temperature dependence of the model parameters (Fig. 2), where the temperature is assumed to be proportional to the non-radiative decay of excitons, $\gamma_{nr}(T)n$. In the analysis of the pulsed laser experiment presented in Fig. 4, we assume that the shift of the exciton resonance, the dephasing rate, and the non-radiative decay rate all depend linearly on the exciton density $n = \langle c^+ c \rangle$. More explicitly, we take $\Delta(n) = \Delta(0) + an$, $\gamma_d(n) = \gamma_d(0) + bn$, and $\gamma_{nr}(n) = \gamma_{nr}(0) + cn$. In both cases, the resulting master equation Eq. (S4) becomes non-linear in $n$ but the steady-state solution may be obtained numerically. As in the linear case, the reflectance is found upon substituting the solution of the master equation back into Eq. (S3).

We note that in the presence of nonlinearities, the master equation may have several distinct steady-state solutions. If more than one solution exists, we choose the relevant one by considering the history of the intensity $I_p$ as illustrated in Fig. 3B.

Dynamics of the optical bistability and simulation of laser heating effect

When the monochromatic CW laser is ~3 nm red detuned from the exciton resonance (Fig. S2A), the reflectance of a MoSe$_2$ monolayer exhibits a sudden jump to a lower value at the laser power $I_p$ ~1.5 mW, and exhibits a prominent hysteresis. When pulsed laser sources with minimum pulse width of tens of nanoseconds are used instead, however, the hysteresis disappears whereas the jump persists. Importantly, the reflectance jumps occur at almost identical $I_p$ values regardless of whether a CW or pulsed laser is used: this is despite the fact that the average power is decreased by as much as a factor of two when a pulsed laser is used. When the MoSe$_2$ monolayer is illuminated with a light pulse with ~1 ns rise/fall time (produced by an electro-optic modulator), the reflectance change occurs in ~3 ns when jumping down to lower reflectance and ~ 1 ns in reverting back (Fig. S2B).

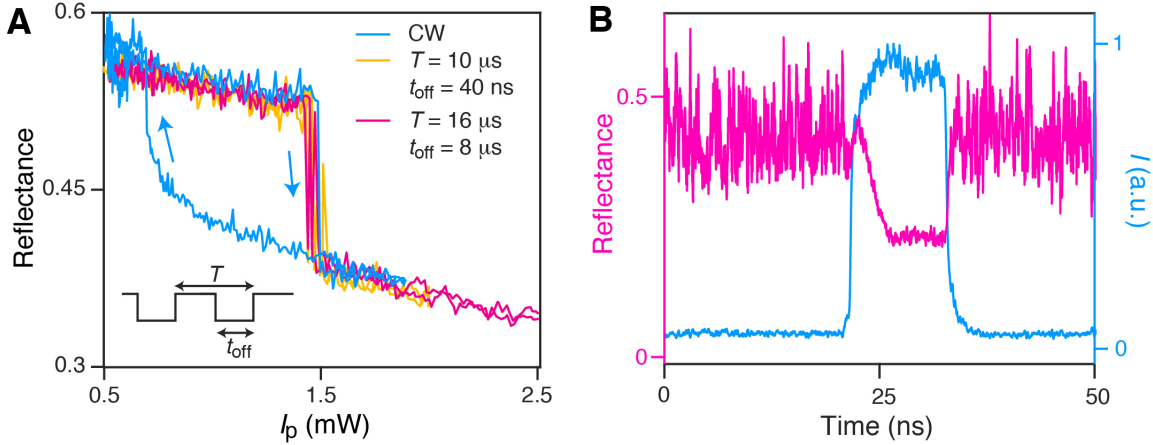

**Figure S2.** (**A**) Optical bistability of MoSe$_2$ reflectance under irradiance of laser red detuned from resonance with excitons. Hysteresis can be observed under CW excitation but not under pulsed excitation. (**B**) Transient response of optical bistability in MoSe$_2$ under a pulse excitation with ~10 ns pulse width.

The nonlinear reflectance predicted by the model described above shows good agreement with the experimental values. Figure S3A shows the increase in the TMD local temperature as a function of incident power for different wavelength, as fitted by the model. To evaluate if the local laser heating can indeed lead to such a temperature increase, we model the heat transfer process of the system with finite element simulations. In the model, the thickness of hBN, MoSe$_2$ and SiO$_2$ layer was taken to match the exact device. The thickness of the Si substrate was set to 5 μm to reduce the mesh size, but the exact thickness does not influence the results. The heat capacity and thermal conductivity values of Si and SiO$_2$ are take from literature. For hBN flakes used in our study with ~100 nm thickness, we assume its heat capacity and thermal conductivity are similar to its bulk values with large anisotropy (*34, 35*). Unfortunately, the heat capacity and thermal conductivity of MoSe$_2$ at low temperatures are unknown. We assume it to be similar to that of BN because of similar crystal structure. Varying the values of MoSe$_2$ heat capacity and thermal conductivity changes the steady state temperature slightly but not significantly. For the heat source, we assume that the absorption only happens within the MoSe$_2$ monolayer, and the laser spot is diffraction limited. Figure S3B shows the simulated stationary-state temperature distribution when the laser power is 1 mW. The finite element simulation using COMSOL produces an increase in local temperature of ~100 K, which agrees well with the modeled values as shown in Fig. S3A.

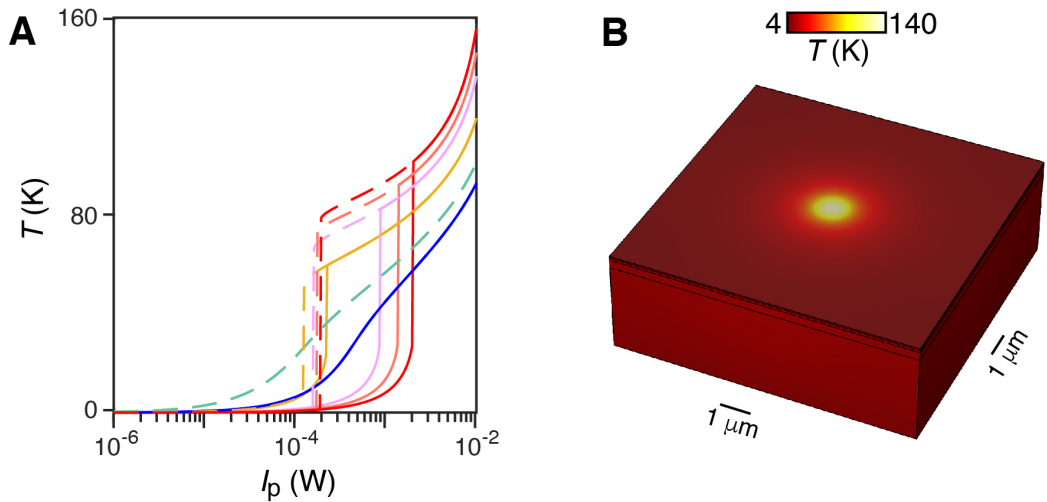

**Figure S3.** (**A**) Modeled local temperature as a function of incident peak power at different detuning. Each wavelength is color-coded in the same way as in Fig. 3. The solid and dashed lines indicate increasing and decreasing power, respectively. (**B**) The local temperature increase simulated by COMSOL with a laser incident power of 1 mW agrees well with the model in (**A**).

Density-dependent resonance shift and decay/dephasing rate for pulsed excitation

The data in Fig. 4B were fitted using a nonlinear model as described in the main text. For the purpose of fitting, it is convenient to express the three original fitting parameters in terms of three new parameters $n_s$, $\theta$ and $\phi$ as $a = \gamma_r n_s \cos\theta$, $b = \gamma_r n_s \sin\theta \cos\phi$, and $c = \gamma_r n_s \sin\theta \sin\phi$. The parameter $\phi$ is constrained to lie between 0 and $\pi/2$ to ensure that the dephasing rate and the non-radiative decay rate increase with density. The fit is then performed by sweeping over different values of $\theta$ and $\phi$ and optimizing $n_s$ at each point. In order to relate the measured data in terms of the laser power to the density $n_s$, we assumed in our calculation that the intensity at the MoSe$_2$ layer is given by the power divided by $\pi\lambda^2$, where $\lambda$ is the wavelength of the incident light. We note that while the quality of the fit depends weakly on the exact value of $\phi$, the fits are much poorer for a density-dependent red shift ($\theta > \pi/2$) than for a blue shift ($\theta < \pi/2$).